\documentclass[twocolumn,a4paper,10pt]{article}
\usepackage{geometry}
\usepackage{amsmath,amssymb}
\usepackage{bm,braket}
\usepackage{multicol}
\usepackage{comment}
\usepackage[dvipdfmx]{graphicx}
\usepackage{cite}
\usepackage{authblk}
\usepackage{nidanfloat}

\title{{\Large Universality Class around the SU(3) Symmetric Point of the Dimer--Trimer Spin-1 Chain}}

\author{Tohru Mashiko, Shunji Moriya, and Kiyohide Nomura}
\affil{Department of Physics, Kyushu University, Fukuoka 819-0395, Japan}
\date{}

\begin{document}

\small
\twocolumn[
\maketitle
\begin{quotation}
We study critical phenomena of the SU(3) symmetric spin-1 chains when adding the SU(3) asymmetric term. To investigate such phenomena, we numerically diagonalize the dimer--trimer (DT) model Hamiltonian around the SU(3) symmetric point, named the pure trimer (PT) point. We analyze our numerical results on the basis of the conformal field theory (CFT). First of  all, we discover soft modes at the wave number $q = 0$ and $q = \pm 2\pi/3$ for the PT point, and then the system is critical. Secondly, we find that the system at the PT point can be described by the CFT with the central charge $c=2$ and the scaling dimension $x=2/3$. Finally, by investigating the eigenvalues of the Hamiltonian in the vicinity of the PT point, we find that there is a phase transition at the PT point from a massive phase to a massless phase. From these numerical results, the phase transition at the PT point belongs to the Berezinskii--Kosterlitz--Thouless (BKT)-like universality class that is explained by the level-1 SU(3) Wess--Zumino--Witten model.\\
\\
\end{quotation}
]

\section{Introduction}
\label{sec:intro}
Recently, there have been major achievements in the development of experiments and quantum simulations of ultracold alkaline earth metallic atoms in an optical lattice\cite{desalvo,gorshkov,taie}. To illustrate these types of materials, the SU($\nu$) symmetric Hubbard model\cite{hubbard} ($\nu$: integer) has especially attracted attention, which is a generalization of the SU(2) Hubbard model. In particular, we focus on the SU(3) symmetric spin-1 chain to which the SU(3) asymmetric term is added. The bilinear-biquadratic (BLBQ) model has the Berezinskii--Kosterlitz--Thouless (BKT)-like transition on the SU(3) symmetric point\cite{itoi}. In contrast, concerning the dimer--trimer (DT) model, there is a study showing that the SU(3) symmetric point is not a phase boundary\cite{oh}. Therefore, we study the DT model around the SU(3) symmetric point in more detail.

The DT model Hamiltonian is defined as
\begin{eqnarray}
	\hat{H}_{\mathrm{DT}}=-\sum_{i=1}^{N}\left[ \cos \theta \hat{D}(i)+\sin \theta \hat{T}(i) \right], \label{dt}
\end{eqnarray}
with competing dimer and trimer interactions. The operators $\hat{D}(i)$ and $\hat{T}(i)$ are defined as follows. To begin with, we let $\hat{\bm{S}}_{i}$ denote the spin-1 operator at site $i$. We then introduce $\hat{\bm{S}}_{ij}\equiv \hat{\bm{S}}_{i}+\hat{\bm{S}}_{j}$ for a pair of adjacent sites ($j \equiv i+1$), and $\hat{\bm{S}}_{ijk}\equiv \hat{\bm{S}}_{i}+\hat{\bm{S}}_{j}+\hat{\bm{S}}_{k}$ for a set of three adjacent sites ($k \equiv i+2$). Then, we define the dimer projection operator $\hat{\mathcal{P}}_{D}(i)$ and the trimer projection operator $\hat{\mathcal{P}}_{T}(i)$ as 
\begin{eqnarray}
	&&\hat{\mathcal{P}}_{D}(i) \equiv \frac{1}{12}\left(\hat{\bm{S}}_{ij}^2-2 \right)\left(\hat{\bm{S}}_{ij}^2-6 \right), \label{dpo}\\
	&&\hat{\mathcal{P}}_{T}(i) \equiv -\frac{1}{144}\left(\hat{\bm{S}}_{ijk}^2-2 \right)\left(\hat{\bm{S}}_{ijk}^2-6 \right)\left(\hat{\bm{S}}_{ijk}^2-12 \right). \label{tpo}
\end{eqnarray}	
Each projection operator gives an eigenvalue +1 for spin singlets, and zero for all other spin multiplets. The operators used in Eq. \eqref{dt} are expressed as
\begin{eqnarray}
	&&\hat{D}(i) \equiv 3\hat{\mathcal{P}}_{D}(i),\\
	&&\hat{T}(i) \equiv 6\hat{\mathcal{P}}_{T}(i).
\end{eqnarray}
The DT model was originally proposed for the sake of explaining the characteristics of a trimer liquid (TL). Oh et al. carried out numerical calculations employing the density-matrix renormalization group (DMRG), and they studied the phases of the DT model according to the parameter $\theta$. The region $\pi/8 <\theta<\pi/4$ is the symmetry-protected topological (SPT) phase. The phase is translationally symmetric and massive. The region $\pi/4<\theta<\pi$, the TL phase, is a massless phase and has soft modes at the wave number $q = 0,\,\pm 2\pi/3$. The point $\theta=\pi/2$, the pure trimer (PT) point, is SU(3) symmetric\cite{oh}. In Ref. \ref{bib:oh}, it was argued that the PT point lies in the TL phase.

Next, we review the BLBQ model around the SU(3) symmetric point to confirm differences and similarities between the two models. The Hamiltonian of the BLBQ model is defined as
\begin{eqnarray}
	\hat{H}_{\mathrm{BLBQ}}=\sum_{i=1}^{N}\left[\cos\theta\left(\hat{\bm{S}}_{i} \cdot \hat{\bm{S}}_{i+1} \right)+\sin\theta\left(\hat{\bm{S}}_{i} \cdot \hat{\bm{S}}_{i+1} \right)^{2} \right]. \label{blbq}     
\end{eqnarray}
The region $-\pi/4<\theta<\pi/4$ is the Haldane phase\cite{haldane}. This phase is translationally invariant and massive\cite{nightin,nomura}. The region $\pi/4<\theta<\pi/2$ is the massless trimerized (spin quadrupolar) phase, which was investigated in several numerical works\cite{fath,schm,lauch}. The massless trimerized phase has soft modes at $q=0,\,\pm 2\pi/3$\cite{fath}. The point $\theta=\pi/4$, which is SU(3) symmetric, is known as the Uimin--Lai--Sutherland (ULS) point\cite{sutherland,uimin,lai,kulish}, which is exactly solvable with the Bethe ansatz. The system at the ULS point is critical, whose universality class is the same as that of the level-1 SU$(3)$ Wess--Zumino--Witten [SU$(3)_{1}$ WZW] model\cite{wess,witten,witten2}. Around the ULS point, numerical studies were carried out\cite{fath,lauch} to calculate the central charge $c$ and the scaling dimension $x$, which determine the universality class of the system. Itoi and Kato analyzed\cite{itoi} systems around the ULS point with the renormalization group (RG) by mapping the ULS model to the general SU$(3)_{1}$ WZW model. They found\cite{itoi} that the phase transition at the ULS point belongs to the BKT-like universality class, which we mention in the next paragraph. 

In a system that belongs to the BKT or BKT-like universality class, the correlation length $\xi$ behaves as\cite{itoi}
\begin{eqnarray}
	\xi \sim
	\begin{cases}
	\exp\left[C(\theta_{C}-\theta)^{-\sigma}\right], \label{corre} \,\,\,\,\,\, (\mathrm{for} \,\, \theta < \theta_{C}) \\
	\infty, \,\,\,\,\,\,\,\,\,\,\,\,\,\,\,\,\,\,\,\,\,\,\,\,\,\,\,\,\,\,\,\,\,\,\,\,\,\,\,\,\,\,\,\,\, (\mathrm{for} \,\, \theta \ge \theta_{C})
	\end{cases}
\end{eqnarray}
where $C$ is a positive constant, $\theta_{C}$ is a phase transition point, and $\sigma$ is a critical exponent. In a system with the U(1) symmetry, such as the 2D classical XY model, it is known that the exponent $\sigma=1/2$ and the central charge $c=1$. This type of phase transition is generally called the BKT transition. Also, in the BKT transition, the spin correlation function decays as\cite{kosterlitz2}
\begin{eqnarray}
	\left \langle \hat{S}^{x}_{i} \hat{S}^{x}_{i+r} \right \rangle = \left \langle \hat{S}^{y}_{i} \hat{S}^{y}_{i+r} \right \rangle \propto r^{-1/4} \left(\ln r \right)^{1/8}. \label{correbkt}
\end{eqnarray}	
Because of the logarithmic correction shown in Eq. \eqref{correbkt}, it has been difficult to calculate critical exponents of the BKT transition by conventional numerical methods. To deal with this bothersome correction, one of the authors developed a method, named level spectroscopy\cite{nomura2}, which is designed to cancel the logarithmic correction by appropriately combining several physical quantities. Therefore, it produces credible results for relatively small systems. 

On the other hand, if the system has symmetries higher than U(1), it can be $\sigma \ne 1/2$ and $c \ne 1$. In this paper, we will call this type of phase transition the BKT-like transition. We deal with the SU$(3)$ BKT-like transition around the PT point on the basis of the theory by Itoi and Kato\cite{itoi}, which can be considered as a generalization of the level spectroscopy (see Appendix).

In this paper, we numerically diagonalize the Hamiltonian of the DT model under periodic boundary conditions (PBC) to investigate critical behaviors near the PT point. The DT model Hamiltonian at the PT point is composed only of exchange operators $P_{ij}$, which is introduced in Sect. \ref{sec:pij}. Numerical results at the PT point are given in Sect. \ref{sec:results} to specify the universality class of the system at the PT point. Numerical results around the PT point are given in Sect. \ref{sec:near} to discuss the phase transition occurring at the PT point. Conclusions and discussions are shown in Sect. \ref{sec:conclusion}. In Appendix, we review the calculations made by Itoi and Kato\cite{itoi}.

\section{Exchange Operator}
\label{sec:pij}
We introduce the exchange operator $\hat{P}_{ii^{\prime}}$, which swaps the spin at site $i$ with that at site $i^{\prime}$, as
\begin{eqnarray}
	\hat{P}_{ii^{\prime}}\ket{\cdots S^{z}_{i} \cdots S^{z}_{i^{\prime}} \cdots} = \ket{\cdots S^{z}_{i^{\prime}}  \cdots S^{z}_{i} \cdots}, \label{pdef}
\end{eqnarray}	
where $\ket{\cdots}$ is a state vector of a spin system and $S^{z}_{i}$ is a spin magnetic quantum number at site $i$. The dimer and trimer projection operators defined in Eqs. \eqref{dpo} and \eqref{tpo} can be rewritten\cite{oh} as
\begin{eqnarray}
	&& \hat{\mathcal{P}}_{D}(i)=\frac{1}{3}\left( \hat{P}_{ij}-\hat{\bm{S}}_{i} \cdot \hat{\bm{S}}_{j} \right), \label{dp} \\
	&& \hat{\mathcal{P}}_{T}(i)=\frac{1}{6}\left( \hat{1}+\hat{P}_{ijk}+\hat{P}^{-1}_{ijk}-\hat{P}_{ij}-\hat{P}_{jk}-\hat{P}_{ki} \right), \label{tp}
\end{eqnarray}
where we define $j$ and $k$ as $j \equiv i+1$ and $k \equiv i+2$, and $\hat{1}$ is the identity operator. The three-site exchange operators\cite{thouless} used in Eq. \eqref{tp} are defined as
\begin{eqnarray}
	&& \hat{P}_{ijk} \equiv \hat{P}_{jk} \hat{P}_{ij} = \hat{P}_{ik} \hat{P}_{jk} = \hat{P}_{ij} \hat{P}_{ik},\\
	&& \hat{P}^{-1}_{ijk} \equiv \hat{P}_{ij} \hat{P}_{jk} = \hat{P}_{jk} \hat{P}_{ik} = \hat{P}_{ik} \hat{P}_{ij}.
\end{eqnarray}

At the PT point, the Hamiltonian is composed only of the exchange operators, which leads to the conservation of the number of spins, $N_{1}$, $N_{0}$, $N_{-1}$ for each state $S^z=1,\,0,\,-1$ respectively. Then, the $3^{N}$ dimensional Hilbert space is reducible to an $\frac{N!}{N_{1}!N_{0}!N_{-1}!}$ dimensional subspace, ($N = N_{1}+N_{0}+N_{-1}$).

\section{The PT Point}
\label{sec:results}

\begin{figure}[b]
\begin{center}
\hspace*{0em}
\includegraphics[width=70mm,height=50mm]{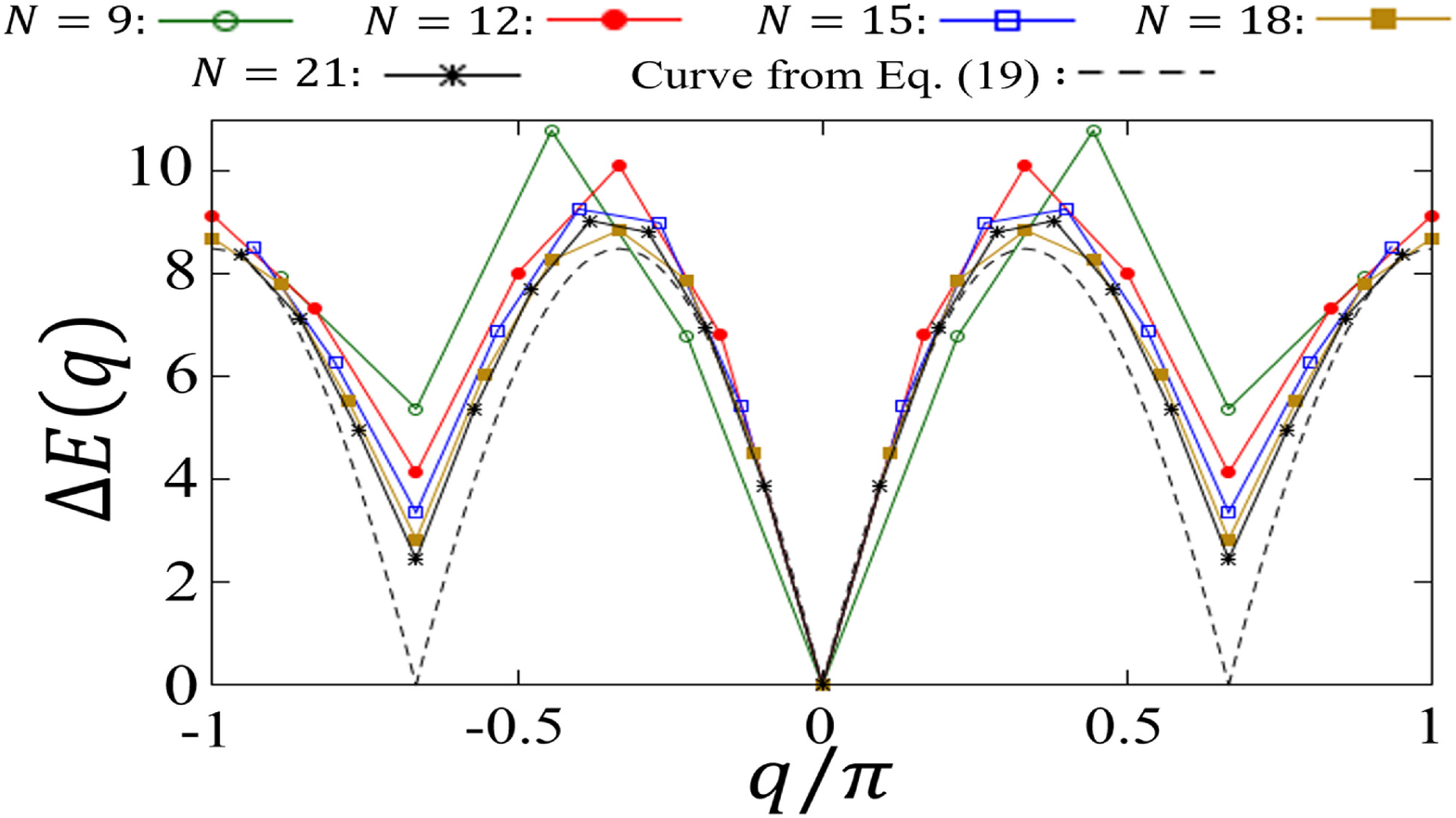}
\end{center}
	\caption{Dispersion curves $\Delta E(q)$ at the PT point for the wave number $q$ with $N=9$--$21$. The dashed line is a curve obtained using Eq. \eqref{eq3}.}
        \label{fig:eq}
\end{figure}

In this section, we show the results of our numerical calculations of the DT model Hamiltonian at the PT point, utilizing the conservation of the number of each spin, $N_{1}$, $N_{0}$, $N_{-1}$, and the translational symmetry. Then, we investigate several physical quantities, namely, the scaling dimension $x$, the central charge $c$, and the coefficients $d$ of the logarithmic correction, to specify the universality class of the system.

First, we let $\hat{T}$ be a translational operator, which shifts spins in the system by one site. $\hat{T}$ has an eigenvalue written as
\begin{eqnarray}
	\hat{T} \ket{\cdots} = \exp \left( iq \right) \ket{\cdots}, \label{qdef}
\end{eqnarray}
where $q$ is the wave number. Under PBC, $\hat{T}^{N}$ is an identity operator. Therefore, the wave number should be $q=2\pi n/N$ ($n$: integer).

The energy eigenvalue $E$ is a function of the wave number $q$ and the total spin quantum number of the system $S_{T}$. Thus, we let $E_{S_{T}}(q)$ denote the lowest energy at certain $q$ and $S_{T}$. We define the difference between $E_{S_{T}}(q)$ and the ground-state energy $E_{g}$ as
\begin{eqnarray}
\Delta E_{S_{T}}(q) \equiv E_{S_{T}}(q) -E_{g}.
\end{eqnarray}
Then, we let $E(q)$ be the lowest energy at a certain $q$ and define the difference between $E(q)$ and $E_{g}$ as
\begin{eqnarray}
        \Delta E(q) \equiv E(q) - E_{g}.
\end{eqnarray}

\subsection{Dispersion curves}
\label{subsec:dis}
Figure \ref{fig:eq} shows dispersion curves $\Delta E(q)$ at the PT point with $N=9$--$21$ as a function of the wave number $q$. We find that the ground-state energy is the lowest energy at $q=0$ and $S_{T}=0$, namely, $E_{g} = E(0) = E_{0}(0)$. Moreover, soft modes appear at $q=0,\,\pm 2\pi/3$ for all the system sizes, as shown in Fig. \ref{fig:eq}. These results are consistent with the theory of Sutherland\cite{sutherland}. In this theory, in the case of the ULS point of the BLBQ model, $\Delta E(q)$ is given by
\begin{eqnarray}
\Delta E(q) = \frac{4\pi}{\sqrt{3}} \left[\cos \left(\frac{\pi}{3} - \left| q \right| \right) - \frac{1}{2} \right],\,\,\,\,\,\,\,\,\,\, \left(0 \le \left| q \right| \le \frac{2\pi}{3} \right) \label{eq}\\ 
\Delta E(q) = \Delta E\left(\left| q \right| - \frac{2\pi}{3} \right),\,\,\,\,\,\,\,\,\,\,\, \left( \frac{2\pi}{3} \le \left| q \right| \le \pi \right) \label{eq2}
\end{eqnarray}
in the thermodynamical limit, $N \rightarrow \infty$. Our numerical results shown in Fig. \ref{fig:eq} also seem to follow
\begin{eqnarray}
	\,\,\,\,\,\,\,\,\, \Delta E(q) = D \left[\cos \left(\frac{\pi}{3} - \left| q \right| \right) - \frac{1}{2} \right], \,\,\,\,\,\,\,\,\,\,\,\,\, \left(0 \le \left| q \right| \le \frac{2\pi}{3} \right) \label{eq3}
\end{eqnarray}
and the same equation as Eq. \eqref{eq2}, where $D$ is a non-universal constant. A dispersion curve gained using Eq. \eqref{eq3} is also shown in Fig. \ref{fig:eq}. We also find that $E(\pm 2 \pi/3) = E_{1}(\pm 2 \pi/3) = E_{2}(\pm 2 \pi/3)$, that is, an eightfold degeneracy. The eightfold degeneracy is composed of the threefold degeneracy of the spin triplet state ($S_{T} = 1$) and the fivefold degeneracy of the spin quintuplet state ($S_{T} = 2$). Considering the fact that soft modes appear at $q=0,\,\pm 2\pi/3$, one should carry out numerical calculations only in cases where $N$ is a multiple of $3$ in later sections as well.

\begin{figure}[t]
\begin{center}
\hspace*{0em}
\includegraphics[width=70mm,height=47.22mm]{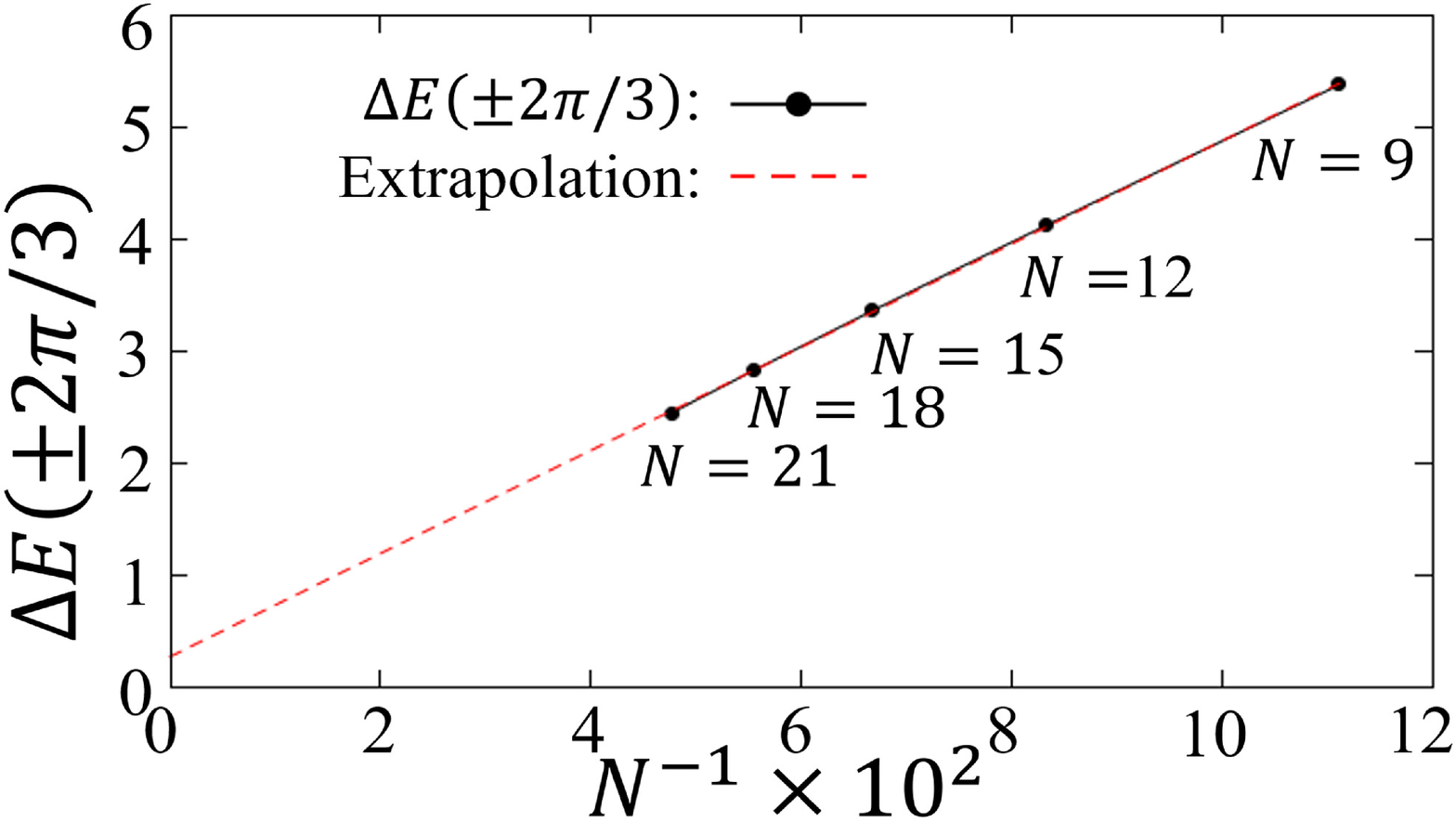}
\end{center}
        \caption{Elementary excitation energy, $\Delta E(\pm 2\pi/3)$, at the PT point as a function of $N^{-1}$.}
        \label{fig:del}
\end{figure}

In Fig. \ref{fig:del}, we replot $\Delta E(\pm 2\pi/3)$ for different system sizes. The excitation energy $\Delta E(\pm 2\pi/3)$ depends linearly on $N^{-1}$. We extrapolate $\Delta E(\pm 2\pi/3)$ with the function $\Delta E(\pm 2\pi/3) = a_{0} + a_{1} N^{-1}$, where $a_{0}$ and $a_{1}$ are constants. We then obtain $a_{0} = 0.27 \pm 0.01$. It seems that a small gap may exist, but this should be massless considering the logarithmic correction, as will be discussed in Sect. \ref{subsec:x}.

\begin{figure}[t]
\begin{center}
\hspace*{0em}
\includegraphics[width=70mm,height=47.22mm]{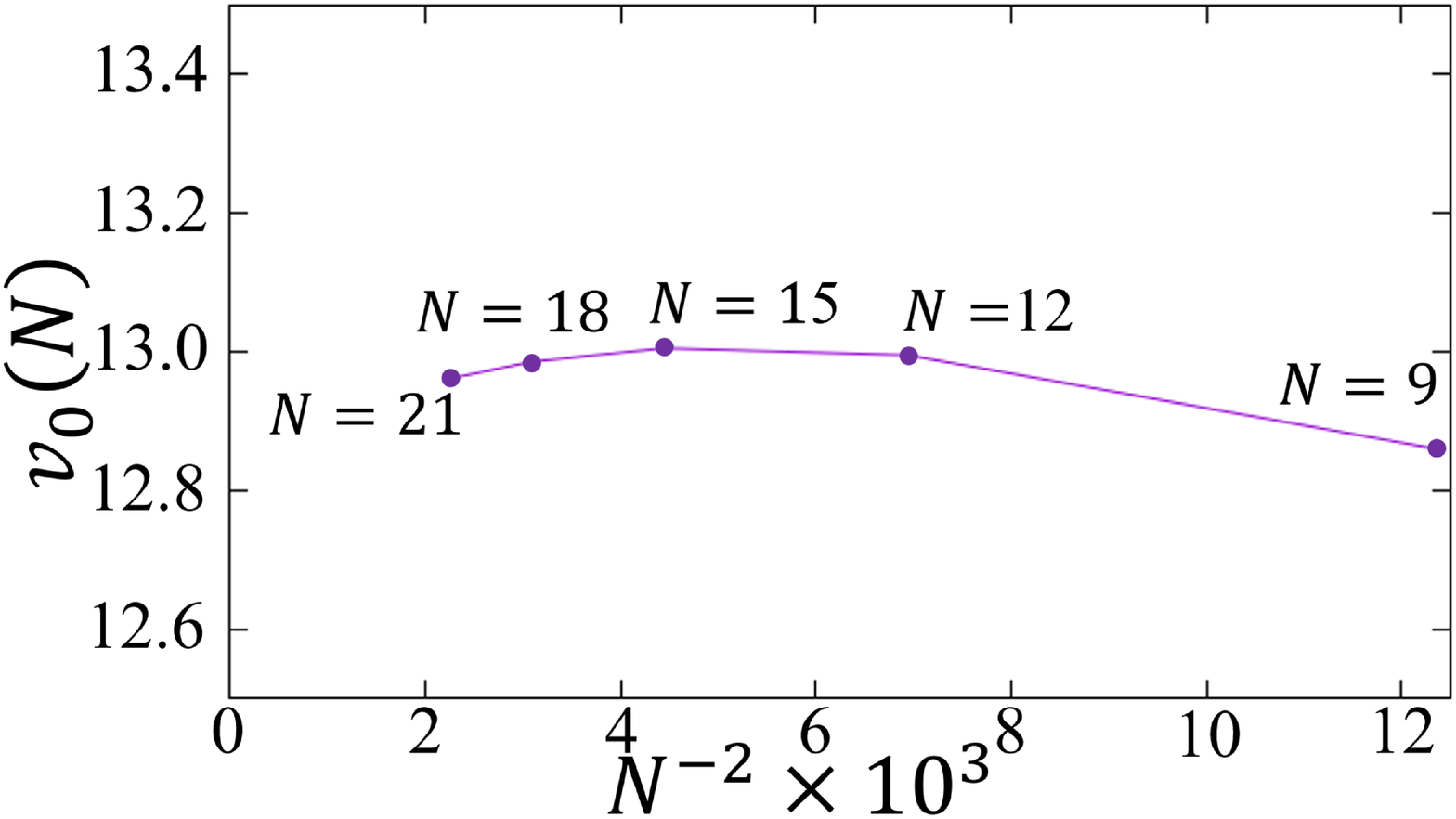}
\end{center}
        \caption{Spin wave velocity with $q=0$ at the PT point as a function of $N^{-2}$.}
        \label{fig:v}
\end{figure}

Additionally, we calculate the spin wave velocity, which is utilized for later calculations of the scaling dimension and central charge. The spin wave velocity $v_{0}$  is defined as
\begin{eqnarray}
	v_{0} \equiv \left. \frac{dE(q)}{dq} \right|_{q=0}. \label{v}
\end{eqnarray}	
The spin wave velocity is a function of $N$, $v_{0}(N)$. In the numerical calculations, we investigate the slope of the spectrum shown in Fig. \ref{fig:eq} to obtain the velocity written as
\begin{eqnarray}
	v_{0}(N)&=&\frac{E(2\pi/N)-E(0)}{2\pi/N}. \label{v0n}
\end{eqnarray}
The values of the velocity are plotted in Fig. \ref{fig:v}.

\subsection{Scaling dimension}
\label{subsec:x}

\begin{table}[b]
        \caption{Values of $x_{S_T}$ and $d_{S_T}$ at the  point illustrated by the SU$(3)_{1}$ WZW model\cite{wess,witten,witten2} corresponding to the line $g_{1}<0,\,g_{2}=0$ in Fig. \ref{fig:bktlike} from Ref. \ref{bib:itoi}.}
\hspace*{3.5em}
\vspace*{0em}
\begin{tabular}{cccc} \hline \hline
& $S_{T}=0$ & $S_{T}=1$  & $S_{T}=2$  \\ \hline
$x_{S_T}$ & $2/3$ & $2/3$ & $2/3$   \\
$d_{S_T}$ & $8/9$ & $-1/9$ & $-1/9$ \\ \hline \hline
\end{tabular}
        \label{tab:co}
\end{table}

In this subsection, we show our numerical results of the scaling dimension. The scaling dimension is one of the critical exponents, which specify a universality class. The elementary excitation energy at a certain $S_T$ follows the equation\cite{itoi,cardy,cardy0.5}
\begin{eqnarray}
	\Delta E_{S_T} \left( \pm \frac{2\pi}{3} \right)=\frac{2\pi v_{0}}{N}\left[x_{S_T}+\frac{d_{S_T}}{\ln (N/N_{0})} \right],\label{de}
\end{eqnarray}
where $x_{S_T}$ is the scaling dimension at $S_{T}$, $d_{S_T}$ is a coefficient depending on $S_{T}$, and $N_{0}$ is a non-universal constant. The $x_{S_T}$ and $d_{S_T}$ take the values\cite{itoi} shown in Table \ref{tab:co}, at the point illustrated by the SU$(3)_{1}$ WZW model corresponding to the line $g_{1}<0,\,g_{2}=0$ in Fig. \ref{fig:bktlike}. Note that the logarithmic correction $[\ln(N/N_{0})]^{-1}$ in Eq. \eqref{de} converges slowly and is about $0.26$ in the case of $N=21$, which is not very small compared with $x=2/3$. Thus, we remove the logarithmic correction in Eq. \eqref{de} using the values in Table \ref{tab:co}, 
\begin{eqnarray}
        \frac{1}{9} \left[\Delta E_{0} \left( \pm \frac{2\pi}{3} \right) + 3\Delta E_{1} \left( \pm \frac{2\pi}{3} \right) + 5 \Delta E_{2} \left( \pm \frac{2\pi}{3} \right) \right] \notag 
	\\= \frac{2\pi v_{0}(N)}{N}x(N), \label{x}
\end{eqnarray}
where we rewrite $v_{0}$ to $v_{0}(N)$ defined in Eq. \eqref{v0n}.

After removing the logarithmic corrections, there remain the correction terms derived from descendant fields of the identity operator with $x=4$\cite{cardy,cardy0.5,rein,kitazawa}. Therefore, the effective scaling dimension $x(N)$ behaves as
\begin{eqnarray}
	x(N) = x + C_{1} N^{-2} +C_{2} N^{-4} + O\left( N^{-6} \right), \label{xf}
\end{eqnarray}
where $C_{1}$ and $C_{2}$ are constants.

\begin{figure}[t]
\begin{center}
\hspace*{0em}
        \includegraphics[width=70mm,height=47.22mm]{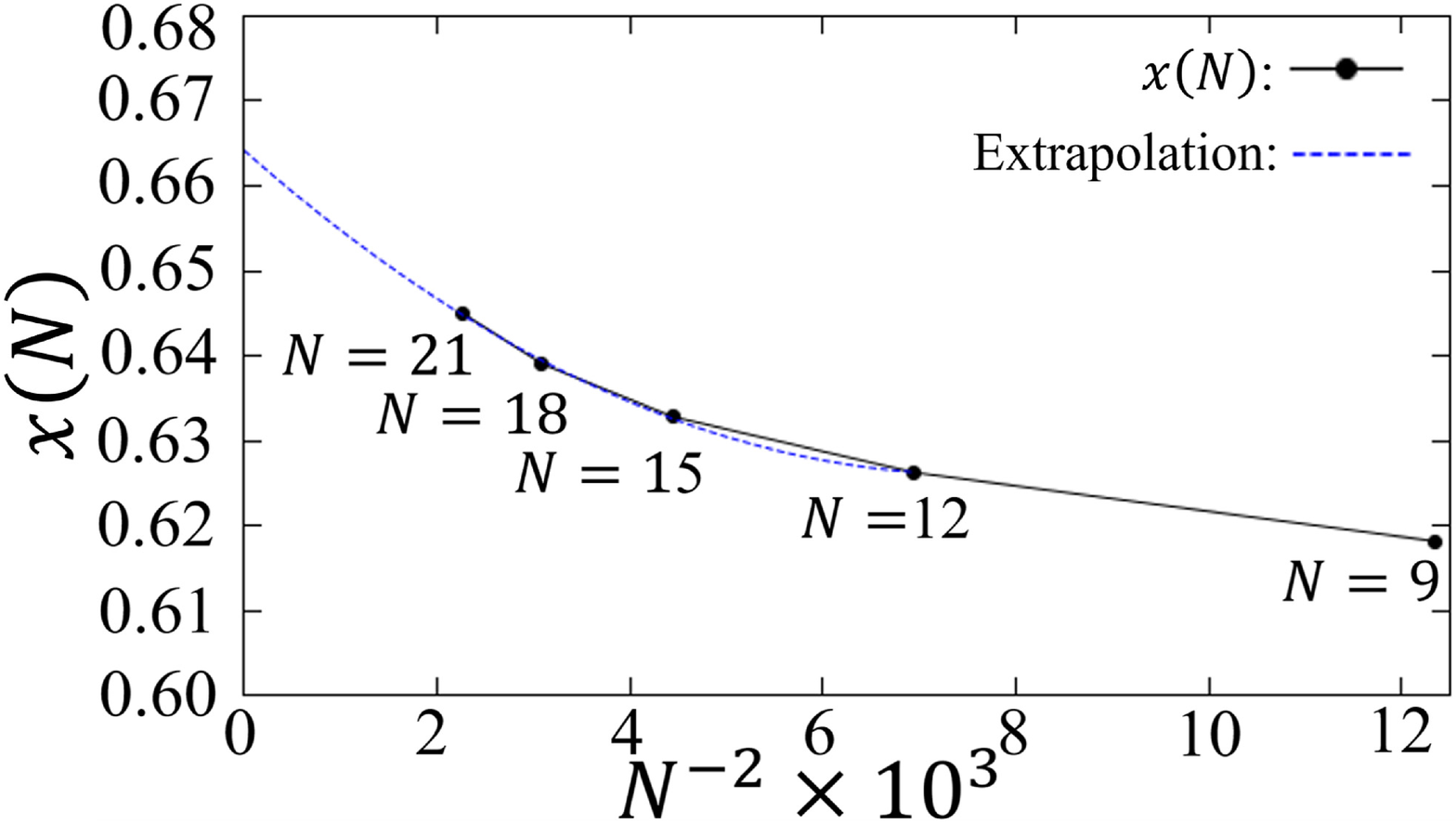}
\end{center}
	\caption{Effective scaling dimension $x(N)$ at the PT point as a function of $N^{-2}$.}
        \label{fig:x}
\end{figure}

Figure \ref{fig:x} shows the numerical results of the effective scaling dimension at the PT point. If we choose a function of the form $x(N)=x+C_{1}N^{-2}+C_{2}N^{-4}$, we obtain $x = 0.6641 \pm 0.0003$ when we extrapolate the $x(N)$ with four points, $N = 12$--$21$. 

These numerical results at the PT point are consistent with the scaling dimension, $x=2/3$, of the SU$(3)_{1}$ WZW model\cite{wess,witten,witten2}.

\begin{figure}[b]
\begin{center}
\hspace*{0em}
\vspace*{0em}
\includegraphics[width=70mm,height=47.22mm]{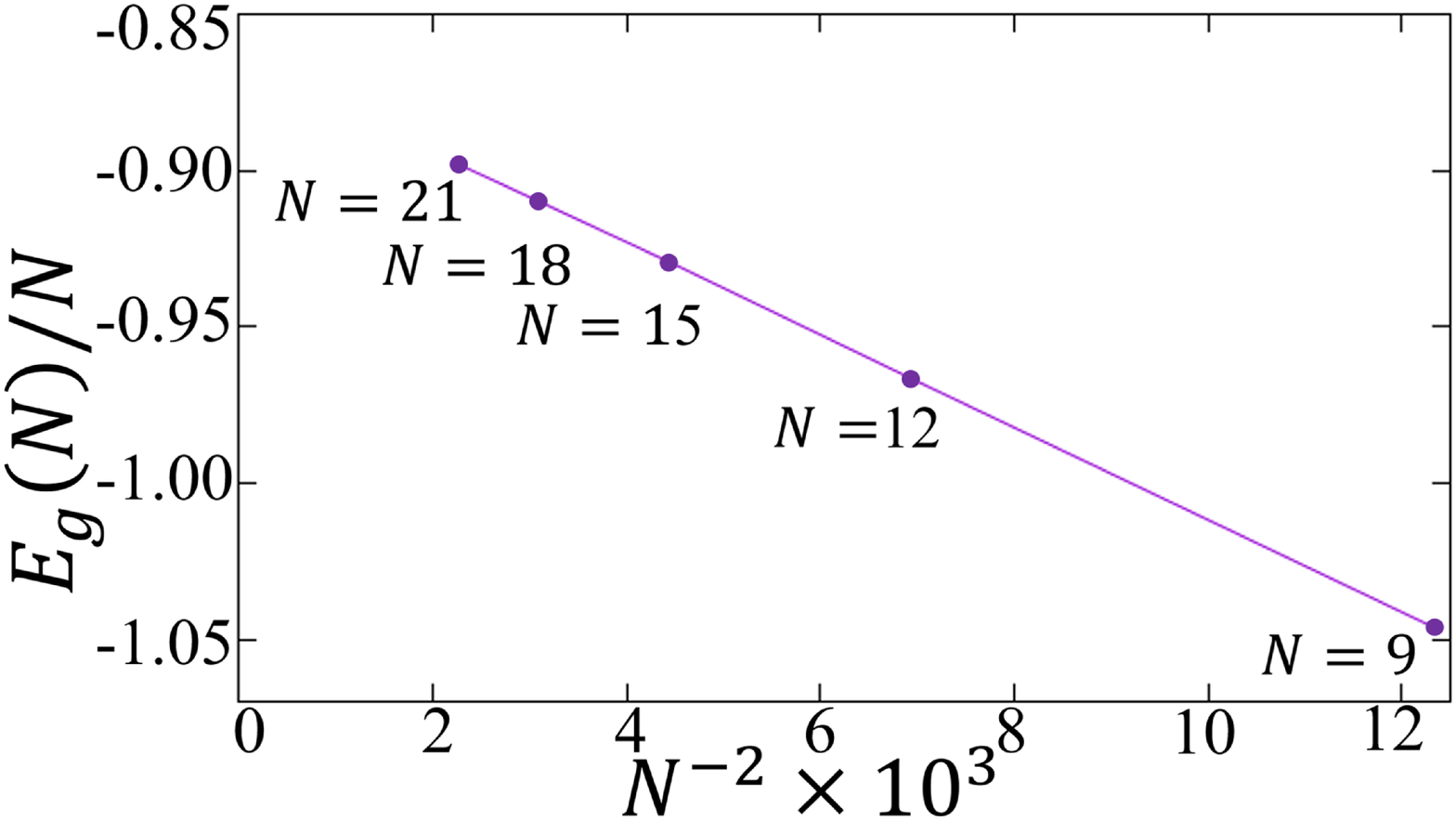}
\end{center}
        \caption{Ground-state energy density $E_{g}/N$ vs $N^{-2}$ at the PT point.}
        \label{fig:eg}
\end{figure}

\begin{figure}[b]
\begin{center}
\hspace*{-0em}
\includegraphics[width=70mm,height=47.22mm]{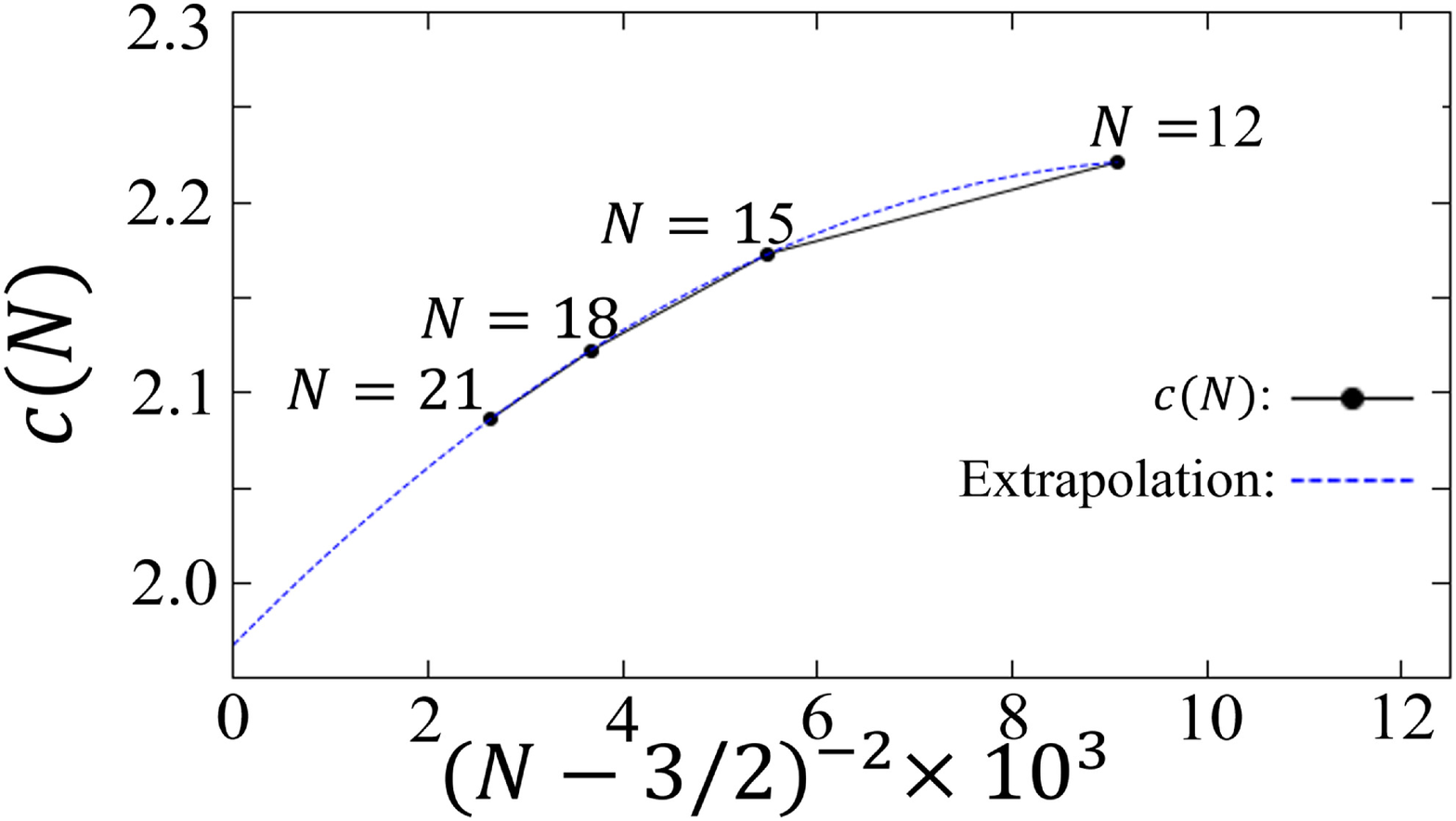}
\end{center}
        \caption{Effective central charge $c(N)$ as a function of $(N-3/2)^{-2}$ at the PT point.}
        \label{fig:c}
\end{figure}

\subsection{Central charge}
\label{subsec:c}
In this subsection, we investigate the central charge, which is also one of the critical exponents, characterizes the quantum anomaly, and specifies the universality class of the system. At the critical point of one-dimensional quantum systems, the ground-state energy density at $N$ should converge\cite{blote,affleck} as
\begin{eqnarray}
        \frac{E_g(N)}{N}=\epsilon_{\infty}-\frac{\pi v_{0} c}{6N^2}, \label{eg}
\end{eqnarray}
where $\epsilon_{\infty}$ is the ground-state energy density in the thermodynamic limit $N \rightarrow \infty$ and $c$  is the central charge. Also, $\epsilon_{\infty}$ and $v_{0}$ are non-universal constants. Note that the central charge has a logarithmic correction\cite{itoi} as a form of $O\left( \left[\ln \left( N/N_{0} \right) \right]^{-3} \right)$ in the $c=2$ CFT. However, since $\left[ \ln(N/N_{0}) \right]^{-3}$ converges much faster than $\left[ \ln(N/N_{0}) \right]^{-1}$, we thus neglect the logarithmic correction in the central charge. The ground-state energy densities are plotted in Fig. \ref{fig:eg} at the PT point. The ground-state energy density depends linearly on $N^{-2}$, consistent with Eq. \eqref{eg}.

In Eq. \eqref{eg}, $E_{g}$ and $v_{0}$ are calculated from the numerical diagonalization and Eq. \eqref{v0n}, but the two constants, $\epsilon_{\infty}$ and $c$, remain as unknown values. Therefore, by removing the constant term $\epsilon_{\infty}$ in Eq. \eqref{eg}, we calculate the effective central charge $c(N)$ as
\begin{eqnarray}
        &&\frac{E_g(N)}{N}-\frac{E_g(N-3)}{N-3} \notag \\
        &&=-\frac{\pi}{6}\left[\frac{v_{0}(N)}{N^2}-\frac{v_{0}(N-3)}{(N-3)^2} \right]c(N). \label{eg3}
\end{eqnarray}

Additionally, similarly to Eq. \eqref{xf}, we extrapolate the effective central charge $c(N)$ as\cite{cardy,cardy0.5,rein,kitazawa}
\begin{eqnarray}
        c(N) = c + D_{1} (N-3/2)^{-2} \,\,\,\,\,\,\,\,\,\,\,\,\,\,\,\,\,\,\,\,\,\,\,\,\,\,\,\,\,\,\,\, \notag \\
\,\,\,\,\,\,\,\,+D_{2} (N-3/2)^{-4} + O\left((N-3/2)^{-6}\right),
\end{eqnarray}
where $D_1$ and $D_2$ are constants.

Figure \ref{fig:c} shows the effective central charge at the PT point for different system sizes. If we choose a function of the form $c(N)=c+D_{1}(N-3/2)^{-2}+D_{2}(N-3/2)^{-4}$ as a fitting function, we obtain $c = 1.9677 \pm 0.0001$ when we extrapolate the $c(N)$ with four points, $N=12$--$21$.

From these results, we conclude that the system at the PT point belongs to the CFT with $c = 2$.

\section{Around the PT Point}
\label{sec:near}
In this section, we investigate the DT model Hamiltonian around the PT point to specify a phase transition and the universality classes of the systems. In our numerical calculations, we make use of the conservation of the magnetization, $M=\sum_{i} S_{i}^{z}$, and the translational symmetry. The reduction of the Hilbert space, mentioned in Sect. \ref{sec:pij}, is not so efficient except at the PT point. Thus, we deal with only smaller systems up to $N=18$.

\begin{figure}[t]
\begin{center}
\hspace*{-0em}
        \includegraphics[width=70mm,height=102.4mm]{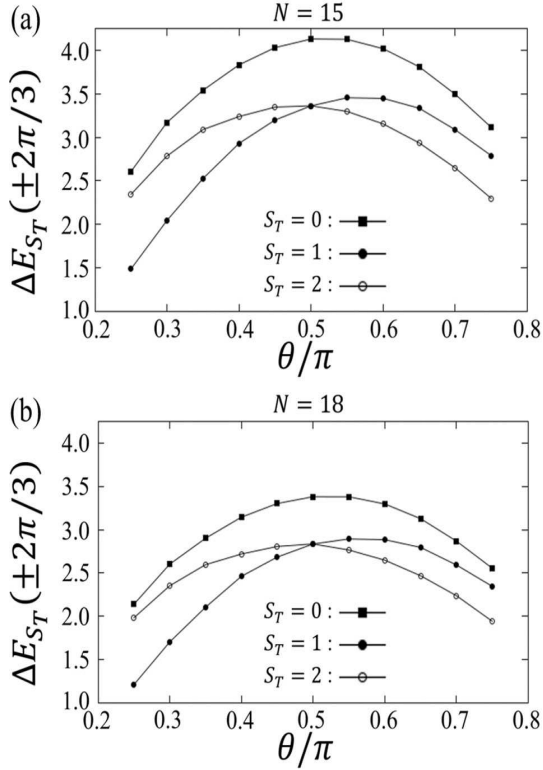}
\end{center}
        \caption{Low-energy spectrum of the DT model with $N=15$ (a) and $N=18$ (b) as a function of $\theta$.}
        \label{fig:de}
\end{figure}

First, we investigate the elementary excitation energies around the PT point. In Fig. \ref{fig:de}, we plot the excitation energy at $q = \pm 2\pi/3$ of the singlet state ($S_{T}=0$), the triplet state ($S_T=1$), and the quintuplet state ($S_T=2$) for various $\theta$ with $N=15$ and $N=18$. As shown in Fig. \ref{fig:de}, $\Delta E_{0}(\pm 2\pi/3)$ is larger than $\Delta E_{1}(\pm 2\pi/3)$ and $\Delta E_{2}(\pm 2\pi/3)$. It can also be seen from Fig. \ref{fig:de} that $\Delta E_{1}(\pm 2\pi/3)$ and $\Delta E_{2}(\pm 2\pi/3)$ are crossing at the PT point. We discuss these numerical results on the basis of the theory of Itoi and Kato\cite{itoi}. Analytically, they studied the action of the fields in the vicinity of the system described by the SU$(3)_{1}$ WZW model\cite{wess,witten,witten2}, as shown in Eq. \eqref{asu}. They derived renormalization-group equations of the action, Eqs. \eqref{rg1} and \eqref{rg2}, and then obtained the trajectories made by the solution of these equations, Eq. \eqref{sol}. As a result of the calculations by the RG method, they found\cite{itoi} that if the system lies in a massless phase corresponding to the second quadrant $g_{1}<0,\,g_{2} \ge 0$ in Fig. \ref{fig:bktlike}, $\Delta E_{S_{T}}(\pm 2\pi/3)$ satisfies the relation 
\begin{eqnarray}
\Delta E_{0}\left( \pm \frac{2\pi}{3} \right) > \Delta E_{1}\left( \pm \frac{2\pi}{3} \right) \ge \Delta E_{2}\left( \pm \frac{2\pi}{3} \right). \label{e012}
\end{eqnarray}
They also found that if the system lies in a massive phase corresponding to the third quadrant $g_{1}<0,\,g_{2}<0$ in Fig. \ref{fig:bktlike}, $\Delta E_{S_{T}}(\pm 2\pi/3)$ satisfies the relation as
\begin{eqnarray}
	\Delta E_{0}\left( \pm \frac{2\pi}{3} \right) > \Delta E_{2}\left( \pm \frac{2\pi}{3} \right) > \Delta E_{1}\left( \pm \frac{2\pi}{3} \right). \label{e021}
\end{eqnarray}
Comparing our numerical results in Fig. \ref{fig:de} with the results obtained on the basis of the theory\cite{itoi}, we find that the region $\theta \ge \pi/2$ in Fig. \ref{fig:de} is a massless phase. We also find that the region $\theta<\pi/2$ in Fig. \ref{fig:de} is a massive phase.

\begin{figure}[t]
\begin{center}
\hspace*{-0em}
        \includegraphics[width=70mm,height=47.22mm]{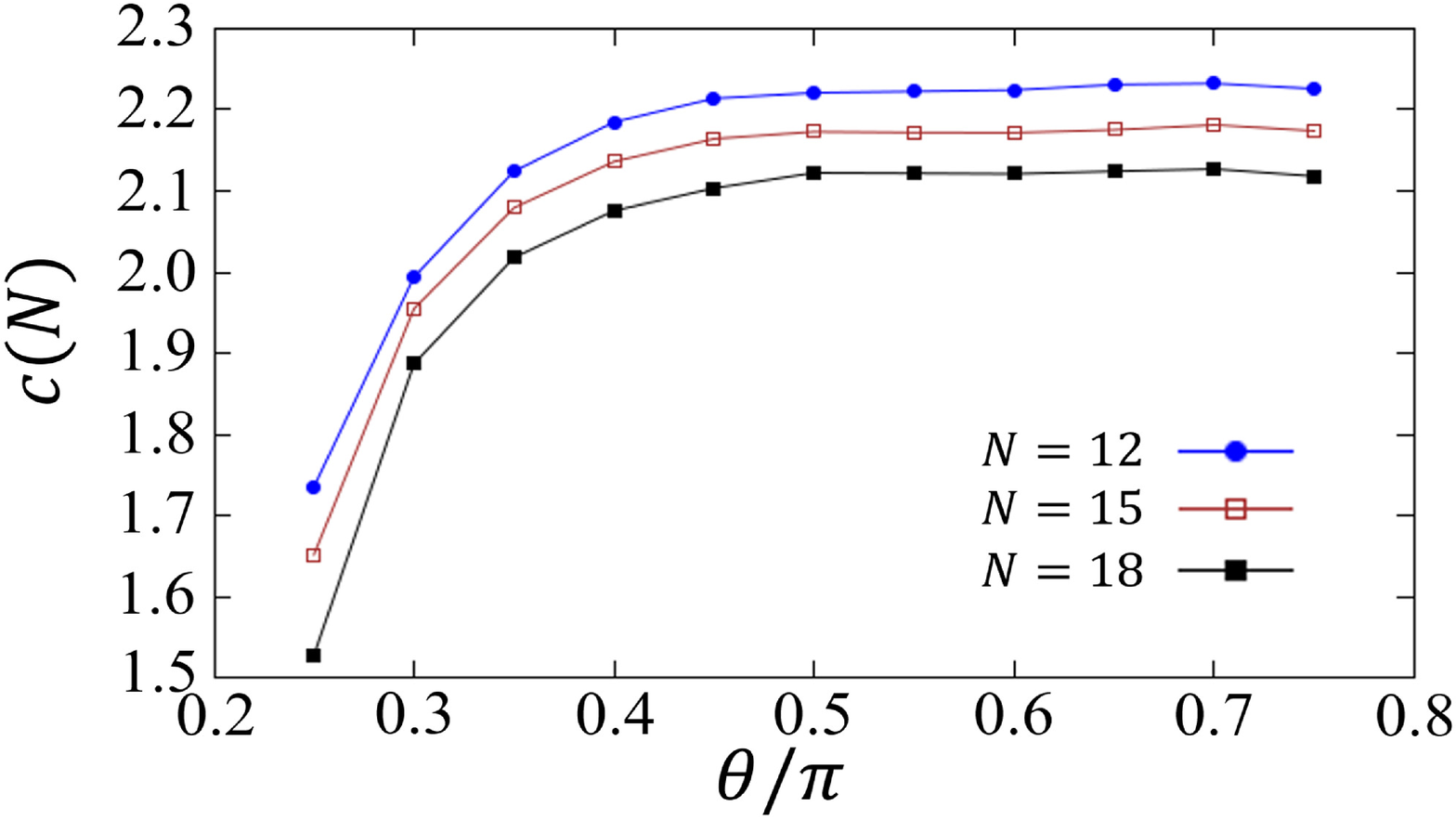}
\end{center}
        \caption{Effective central charge gained by the method shown in Eq. \eqref{eg3} as a function of $\theta$.}
        \label{fig:cth}
\end{figure}

To investigate the universality classes of the systems around the PT point, we plot the effective central charge as a function of $\theta$ in Fig. \ref{fig:cth} with $N=12$--$18$. The effective central charge was firstly investigated with numerical calculations in the case of the CFT with $c=1$\cite{okamoto}. In this study, we calculate the effective central charge using Eq. \eqref{eg3}. Although Eq. \eqref{eg} is true only in the case of the critical phase (or massless phase), we can apply Eq. \eqref{eg3} even to systems in a massive phase. We find that the effective central charge shows a sharp decline in the region $\theta < \pi/2$ in all cases of $N=12$--$18$. In contrast, in the region $\theta \ge \pi/2$, the effective central charges smoothly converge to $c=2$ as $N \rightarrow \infty$ (also see Fig. \ref{fig:c}). These results are consistent with Zamolodchikov's $c$-theorem\cite{zamolodchikov}. 

In summary, from Figs. \ref{fig:de} and \ref{fig:cth} with the CFT plus the RG\cite{itoi,zamolodchikov}, we conclude that the region $\theta \ge \pi/2$ is described by the $c=2$ CFT (massless phase), whereas the region $\theta < \pi/2$ is a massive phase. In addition, from the theory of Itoi and Kato\cite{itoi}, one can say that the scaling dimension is $x = 2/3$ in the $c=2$ CFT phase ($\theta \ge \pi/2$). The SU$(3)_{1}$ BKT-like transition occurs at the PT point.

\section{Conclusions}
\label{sec:conclusion}
We have investigated the DT model to clarify the critical behavior around the PT point, by numerically diagonalizing the DT model Hamiltonian. First of all, we find that soft modes appear at the wave number $q=0,\,\pm2\pi/3$ for the PT point, and the system is critical. Secondly, the PT point can be described by the CFT with $c=2$ and $x=2/3$, or more precisely, the SU$(3)_{1}$ WZW universality class\cite{wess,witten,witten2}. Thirdly, there occurs a phase transition at the PT point from a massive phase to a massless phase. 

As for the spin correlation function at the PT point, it is expected\cite{itoi} to be
\begin{eqnarray}
        \left \langle \hat{\bm{S}}_{i} \cdot \hat{\bm{S}}_{i+r} \right \rangle \propto \cos \left(\frac{2\pi}{3}r \right) r^{-4/3} (\ln r)^{2/9}, \label{correbktlike}
\end{eqnarray}
from Table \ref{tab:co} and Fig. \ref{fig:x} ($x=2/3$, $d=-1/9$). The spin-quadrupolar correlation function is also expected\cite{lauch,stou} to be
\begin{eqnarray}
	&&	\left \langle \hat{Q}^{\mu \nu}_{(i)} \hat{Q}_{(i+r) \mu \nu} \right \rangle \propto \cos \left(\frac{2\pi}{3}r \right) r^{-4/3} (\ln r)^{2/9}, \label{quacorrebktlike} \\
	&& \hat{Q}^{\mu \nu}_{(i)} \equiv \frac{1}{2} \{ \hat{S}_{i}^{\mu}, \hat{S}_{i}^{\nu} \} - \frac{2}{3} \delta^{\mu\nu}, \notag 
\end{eqnarray}
from Table \ref{tab:co} and Fig. \ref{fig:x}. Here, $\hat{Q}^{\mu \nu}_{(i)}$ is the spin-quadrupolar order parameter at site $i$, which is symmetric and traceless. In the region $\theta > \pi/2$ of the DT model, it is expected\cite{lauch,itoi} that the spin-quadrupolar correlation is more dominant than the spin correlation. As for the critical exponent $\sigma$ defined in Eq. \eqref{corre}, it should be $\sigma=3/5$\cite{itoi}.

As mentioned in Sect. \ref{sec:intro}, Oh et al. argued\cite{oh} that the PT point is not the phase transition point. The discrepancy between our results and those of Oh et al. firstly comes from the fact that Oh et al. did not consider\cite{oh} the logarithmic correction shown in Eqs. \eqref{correbkt}, \eqref{correbktlike}, or \eqref{quacorrebktlike}. Even in large systems with $N\approx 10000$, the logarithmic correction is not small\cite{nomura3}, and thus, wrong conclusions often follow from simply utilizing the DMRG without considering the logarithmic correction. Secondly, since they did not calculate critical exponents ($x$ and $c$)\cite{oh}, one cannot trace their reasoning on how to determine the TL phase boundary.

On the other hand, we carry out the calculations with the numerical diagonalization under PBC. By combining the CFT with the finite size scaling, similarly to the level spectroscopy, one can calculate critical exponents (see Figs. \ref{fig:x}, \ref{fig:c}, \ref{fig:de}, and \ref{fig:cth}). Critical exponents from our numerical data and those from the theory\cite{itoi} are consistent within numerical errors. Therefore, we conclude that the TL phase boundary is located at the PT point.

We also believe that our numerical results can be applied to experiments and quantum simulations explained by the SU($\nu$) symmetric Hubbard model\cite{hubbard} described in Sect. \ref{sec:intro}.

\section*{Acknowledgements}
We are grateful to H. Katsura for very constructive discussions and comments on our work. We also thank J. Fukuda for carefully reading the manuscript and giving useful advice for corrections. C. Itoi gave us very fruitful advice about the perturbations around the SU(3) symmetric systems.

\section*{Appendix}
\label{sec:app}
In this section, we review the RG calculation by Itoi and Kato\cite{itoi} to investigate the critical behavior around the system of spin chains illustrated by the SU$(\nu)_{1}$ WZW model\cite{wess,witten,witten2}.

First, we let $x_{0}$ be the time in the system and $x_{1}$ be the position of the field. We then put $z$ and $\bar{z}$ as
\begin{eqnarray}
z \equiv x_{0} + i x_{1}, \,\,\,\,\,\, \bar{z} \equiv x_{0} - i x_{1}.
\end{eqnarray}
We define the action $\hat{\mathcal{A}}$ as
\begin{eqnarray}
	\hat{\mathcal{A}} \equiv \hat{\mathcal{A}}_{\mathrm{SU}(\nu)_{1}} + \sum_{i=1}^{2} g_{i} \int \frac{d^{2}z}{2\pi} \hat{\Phi}^{(i)} \left( z , \bar{z} \right), \label{asu}
\end{eqnarray}
where $\hat{\mathcal{A}}_{\mathrm{SU}(\nu)_{1}}$ is the action of the free fields in the SU$(\nu)_{1}$ WZW model\cite{wess,witten,witten2}. Both $\hat{\Phi}^{(1)}$ and $\hat{\Phi}^{(2)}$ are operators of the marginal or relevant field with rotational symmetry and chiral $\mathbb{Z}_{\nu}$ symmetry. In particular, $\hat{\Phi}^{(1)}$ is SU$(\nu)$ symmetric, and $\hat{\Phi}^{(2)}$ is SU$(\nu)$ asymmetric but O($\nu$) symmetric. The scaling variables $g_{1}$ and $g_{2}$ are perturbational parameters. If $g_{2} = 0$, the system remains SU$(\nu)$ symmetric regardless of the value of $g_{1}$. If $g_{2} \ne 0$, the SU$(\nu)$ symmetry of the system is broken. According to Itoi and Kato\cite{itoi}, the renormalization-group equations for the scaling variables become
\begin{eqnarray}
	\frac{dg_{1}}{dl} &=& \frac{1}{\sqrt{{\nu}^{2}-1}} \left( \nu g_{1}^{2} + 2 g_{1} g_{2} \right), \label{rg1} \\
	\frac{dg_{2}}{dl} &=& - \frac{1}{\sqrt{{\nu}^{2}-1}} \left( \nu g_{2}^{2} + 2 g_{1} g_{2} \right), \label{rg2} \\
	l &\equiv& \ln (N/N_{0}). \notag
\end{eqnarray}
From Eqs. \eqref{rg1} and \eqref{rg2}, there is a fixed point at $g_{1}=g_{2}=0$. Moreover, in the case of $g_{1}=0$, it remains $0$ after the renormalization, and $g_{2}$ diverges or converges as
\begin{eqnarray*}
(g_{1},g_{2}) \rightarrow 
\begin{cases}
	(0,0),\,\,\,\,\,\,\,\,\,\,\,\,\,\, (\mathrm{for}\,\,g_{2}>0) \\
	(0,-\infty).\,\,\,\,\,\,\,\,\, (\mathrm{for}\,\,g_{2}<0)
\end{cases}
\end{eqnarray*}
Also, in the case of $g_{2}=0$, it remains $0$ after the renormalization, and $g_{1}$ diverges or converges as
\begin{eqnarray*}
(g_{1},g_{2}) \rightarrow 
\begin{cases}
5, 1980	(\infty,0),\,\,\,\,\,\,\,\,\,\,\,\, (\mathrm{for}\,\,g_{1}>0) \\
	(0,0).\,\,\,\,\,\,\,\,\,\,\,\,\,\,\, (\mathrm{for}\,\,g_{1}<0)
\end{cases}
\end{eqnarray*}

\begin{figure}[t]
\begin{center}
\hspace*{-0em}
        \includegraphics[width=90mm,height=53.96mm]{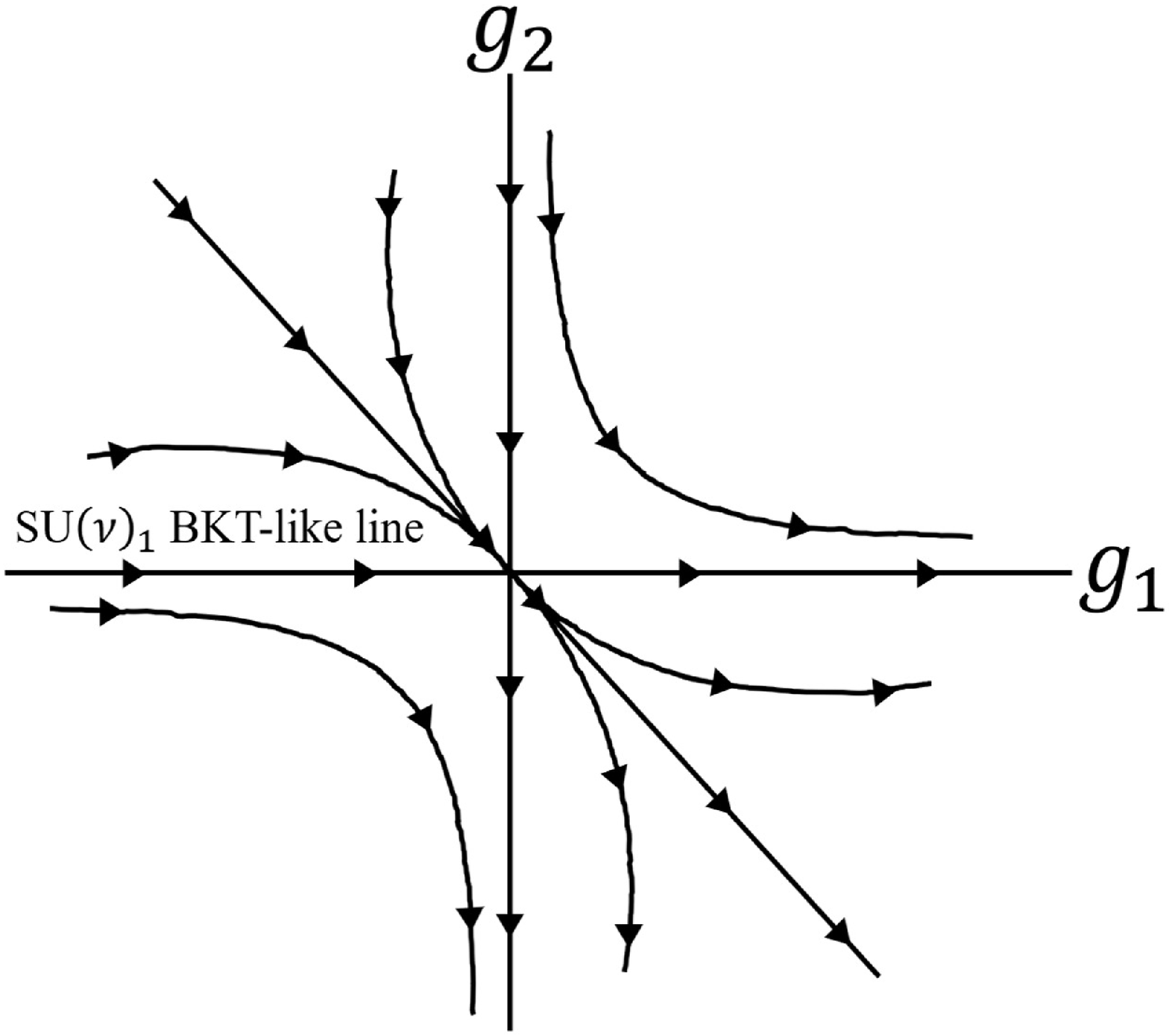}
\end{center}
        \caption{Trajectory gained from the solution of renormalization-group equations, Eq. \eqref{sol} in the case of $\nu>2$.}
        \label{fig:bktlike}
\end{figure}

Here, we  put $X \equiv g_{1} - g_{2}$ and $Y \equiv -g_{1} - g_{2}$. From Eqs. \eqref{rg1} and \eqref{rg2}, we obtain two equations as
\begin{eqnarray}
	\frac{d}{dl} \left( X^{2} - Y^{2} \right) &=& \frac{\nu-2}{\sqrt{{\nu}^{2}-1}} \left( X^{2} - Y^{2} \right) X, \label{rg3} \\
	\frac{d}{dl} \left| Y \right| &=& - \frac{\nu}{\sqrt{{\nu}^{2}-1}} \left| Y \right| X. \label{rg4}
\end{eqnarray}
Using these equations, we obtain
\begin{eqnarray}
	\frac{d}{dl} \left[ \frac{X^{2} - Y^{2}}{\left| Y \right|^{(\nu-2)/\nu}} \right] = 0. \label{rg5}
\end{eqnarray}
In conclusion, a solution of the renormalization-group equations is found to be
\begin{eqnarray}
	X^{2} - Y^{2} = C \left| Y \right|^{(\nu-2)/\nu}, \label{sol}
\end{eqnarray}
where $C$ is a constant.

From Eq. \eqref{sol}, flows\cite{itoi} of the RG can be drawn, as shown in Fig. \ref{fig:bktlike}. Then, we can discuss critical behaviors around the system described by the SU$(\nu)_{1}$ WZW model\cite{wess,witten,witten2}, by analyzing the convergence and divergence of the perturbational parameters, $g_{1}$ and $g_{2}$. The graph can be divided into six regions according to the values of $g_{1}$ and $g_{2}$. The parameters diverge or converge differently depending on the regions they belong to. After repeating the renormalization infinite times, they diverge or converge as
\begin{eqnarray*}
(g_{1},g_{2}) \rightarrow 
\begin{cases}
	(\infty,0),\,\,\,\,\,\,\,\,\,\,\,\, (\mathrm{for}\,\,g_{1}>0,\,g_{2}>0) \\
	(0,0),\,\,\,\,\,\,\,\,\,\,\,\,\,\, (\mathrm{for}\,\,g_{1}<0,\,g_{2}>0) \\
	(0,-\infty),\,\,\,\,\,\,\,\, (\mathrm{for}\,\,g_{1}<0,\,g_{2}<0) \\
	(\infty,0),\,\,\,\,\,\,\,\,\,\,\,\, (\mathrm{for}\,\,g_{1}>0,\,g_{2}<0,\,g_{1}+g_{2}>0) \\
	(0,-\infty),\,\,\,\,\,\,\,\, (\mathrm{for}\,\,g_{1}>0,\,g_{2}<0,\,g_{1}+g_{2}<0) \\
	(\infty,-\infty).\,\,\,\,\,\,\, (\mathrm{for}\,\,g_{1}>0,\,g_{2}<0,\,g_{1}+g_{2}=0) \\
\end{cases}
\end{eqnarray*}
Therefore, the region $g_{1}<0,\,g_{2}>0$ corresponds to a massless phase, and the other regions correspond to different massive phases. 

This theory by Itoi and Kato\cite{itoi} is a generalization of the level spectroscopy\cite{nomura2} of the BKT transition, which is equivalent to the case of $\nu=2$ of the theory by Itoi and Kato\cite{itoi}. In the case of $\nu = 3$, fields in the theory\cite{itoi} correspond to the systems of the DT model and the BLBQ model. The transition point that we deal with in this paper corresponds to the line of $g_{1}<0,\,g_{2}=0$, named the SU$(3)_{1}$ BKT-like line. According to Itoi and Kato\cite{itoi}, the relation Eq. \eqref{e012} holds only in the case where the parameters belong to the second quadrant, $g_{1}<0,\,g_{2}>0$. Also, according to Fig. \ref{fig:bktlike}, the region $\theta < \pi/2$ in Figs. \ref{fig:de} and \ref{fig:cth} corresponds to the case where the parameters belong to the third quadrant, $g_{1}<0,\,g_{2}<0$. That is, the BKT-like transition in this paper corresponds to the transition between the second quadrant and the third quadrant.

\end{document}